\documentclass[paper]{ieice}
\usepackage[pdftex]{graphicx,xcolor}
\usepackage[fleqn]{amsmath}
\usepackage{newtxtext}
\usepackage[varg]{newtxmath}
\usepackage[hyphens]{url}
\usepackage{multirow}
\usepackage{xurl}

\field{}
\title{Examining the Factors of Place Sameness: A Classroom Re-creation Task in a Virtual Environment}
\authorlist{%
 \authorentry[aoyagi12@komazawa-u.ac.jp]{Saizo Aoyagi}{m}{KomazawaU}\MembershipNumber{xxxx}
 \authorentry{Satoshi Fukumori}{n}{KagawaU}
 \authorentry{Kenji Hirose}{n}{HokkaidoU}
 \authorentry{Takayoshi Kitamura}{n}{KagawaU}
}
\affiliate[KomazawaU]{The author is with the Faculty of Global Media Studies, Komazawa University, Setagaya-ku, 154-8525 Japan}
\affiliate[KagawaU]{The authors are with the Faculty of Engineering and Design, Kagawa University, Takamastu-shi, 2217-20 Japan}
\affiliate[HokkaidoU]{The author is with the Center for Human Nature, Artificial Intelligence, and Neuroscience, Hokkaido University, Sapporo-shi, 060-0812 Japan}

\received{2015}{1}{1}
\revised{2015}{1}{1}

\begin{document}
\maketitle
\begin{summary}
Virtual re-creations of real-world places are attractive and becoming more popular.
Attractiveness of re-created place is commonly determined by the degree of visual similarity to the original places.
Even if the re-creation has poor similarity to the original, there are cases in which people recognize and allow the re-creation to have similarity as the same entity. 
However, the mechanisms and factors behind this contradiction have not yet been studied.
This is the first study focusing on the activity and meaning of people when they feel sameness to a re-created place.
This paper investigates the concept of place sameness, which is defined as the degree of similarity between an original and a virtual re-creation that represents the original place.
Further, the factors of place sameness are explored using a classroom re-creation task with virtual reality (VR) technology. 
The participants were instructed to recreate two classrooms within virtual environments using virtual fixtures. 
As results, the display devices in Room B, the student desks in Room A at the memory index, display devices in Room A, and student desks and others in Room B at the average distance had moderate or large correlations with the place sameness index.
The results suggest that the reproducibility of the number of objects related to activities in a place, and the inaccuracy of the positions of objects are factors of place sameness. 
Our own interpretation of the uncanny valley effect of a place was also partially observed.
The main contributions of this study are the proposal of the concept of place sameness as a new perspective for virtual re-creation research and the finding of promising factors for that.
\end{summary}

\section{Introduction}

In recent years, several studies and entertainment contents have implemented three-dimensional computer graphics (3DCG) to replicate real-world places in virtual environments. 
For example, the Notre-Dame Cathedral \cite{NotreDame} was re-created by Ubisoft Games Inc., and Shibuya \cite{VirtualShibuya} was re-created using an online event platform Cluster\footnote{https://cluster.mu/}.

Improvements in 3DCG technology, restrictions on travel and outdoor activities during the pandemic, and the collapse of historic buildings have led to a rapid increase in virtual re-creations.
A virtual re-creation has a unique charm and it attracts people. 
Ubisoft Games Inc. received significant media attention for launching a virtual tour of the Notre Dame Cathedral after its collapse\cite{NotereDameNews}. 
The Halloween festival on Virtual Shibuya is one of the most popular events hosted on the platform\cite{VShibuyaNews}. Even three years after its launch, Virtual Shibuya continues to host live music and concerts.

In spite of its attractiveness, there is no clarity on what makes virtual re-creations of places so attractive.
The essence of re-creation is to create again something else that already exists, this could be the source of its unique charm.
If the mechanism and source of this charm can be identified, it could enable us to make more attractive place re-creations.
Meanwhile, this examination raises philosophical and hypothetical questions such as 
1)	Can we achieve a perfect match between a virtual and a physical place? 
2)	Is it feasible to transfer a destroyed physical building to the virtual world without reconstruction in the real world?
3)	What does it mean to be the same?

In this study, our first step toward addressing these questions is to identify factors that increase the similarly between virtual and physical places.
We focus on two examples of traditional Japanese places in an attempt to improve our understanding of factors of similarity. 

Ise Jingu is a historic shrine in Japan that has a rebuilding ceremony.
The shrine is rebuild every 20 years in a ceremony called ``Shikinen-Zotai.''
Although they are commonly perceived to have remained unchanged throughout history, their appearance and structure are changed in the past rebuilding ceremonies\cite{Shimizu2006}. 
Despite having a new architectural structure, the changed Ise Jingu is recognized as the same entity.
This recognition may seem contradictory; however, it has consistency.
One reason for this consistency is that the object of worship is believed to contain the spirit of a deity (Goshintai), remains the same. 
Another reason is the belief that the activity of rebuilding ceremonies continues uninterruptedly\cite{Lopes2007}.
Thus, Ise Jingu is re-created based on the interaction between the place and human's spiritual and physical activities.

During the Edo period in Japan, replica buildings called ``Utsushi'' were believed to inherit the spirit of the original buildings, regardless of any differences in their architectural structures\cite{Kaneko2006}.
The Sanjusangen-do temple in Fukagawa is one such popular instance of Utsushi from the Edo period.
This was a replica of the Sanjusangen-do temple in Kyoto. 
Although these had many differences in the architectural structure and fixture, people assigned the same meaning to both Fukagawais's and Kyoto's temples. 
The meaning was "a place where people can shoot a bow in its garden as  an archery area".
Therefore, the temple was re-created for performing the same human activity in the place.

In the two examples, people believe that the place has ontological equivalences,
even though imperfect visual similarities and material equivalence  
with the original places as entities
%
What are the essential factors of equivalence?
According to humanistic geography\cite{Relph1976}, a place includes not only a collection of physical objects and areas of space but also subjective meanings that people assign to it and social activities that take place there.
In other words, a place consists of three components, i.e., space, meaning, and activity.
Utsushi in the case of the Sanjusangen-do temple provides a space for activity, which is to shoot a bow.
Shikinen-Zotai in Ise jingu provides a space for meaning and activity to continue worship.

The meanings and activities can be represented to be a part of, or inside, architectural structures and fixtures in the re-creations in the real world.
Virtual re-creations may also have ontological equivalence by representing meaning and activity as a part of, or inside, architectural structures and fixtures.
However, no study of factors of equivalence to the original place has yet addressed how activity and meaning expressed onto an virtual object re-created.

In this study, the authors explore the architectural structures and fixtures and identify how they can be re-created in the virtual place to represent meanings and activities of the real place.
To this end, a feeling of equivalence between a virtual re-created place and the original place is conceptualized as place sameness.
Next, virtual re-creations of university classrooms are created and 
an experiment with student participants is conducted to explore factors of place sameness focusing on the type of architectural structures and fixtures and what activity and meaning their arrangement reflects.
Finally, hypotheses about place sameness factors are presented based on the experimental results.


Existing studies of similarity or identity\cite{Kahneman1992}\cite{Lopes2007}\cite{Baccini2018} have not dealt with virtual places, and studies of virtual worlds\cite{NagakamiTakeuchi2008} have not done similarity or identity issues of places.
The main contribution of this study is the proposal of the concept of place sameness in the research field of similarity or identity issues of places.
Application of the experimental approach utilizing a classroom re-creation task is also a contribution to the field, which is predominantly case studies.

In Section 2, we describe related studies and show the significance of this study compared to them.
In Section 3, the concept of place sameness is proposed.
In Section 4, we describe the method, the results, discussion of the experiment using the classroom re-creation task.
In Section 5, we state our conclusions.

\section{Related Studies}
\subsection{Sameness}

To the best of our knowledge, this is the first study to address ontological equivalence or sameness between real and virtual places.
Perceptual psychology investigated conditions under which objects are perceived to be the same, even if there are visual differences \cite{Kahneman1992}. Their research targets objects smaller than a person, not places surrounding the person.

The philosophy and aesthetics of architecture suggest that logical conditions for the identity of a building, including its ability to persist as the same entity despite spatial relocation, restoration, or reconstruction, have been studied for the periodic reconstructions of Ise Jingu\cite{Lopes2007}, similarity to music performance\cite{Wicks1994}, and similarity between cooking and architecture\cite{Baccini2018}
However, these studies are case studies and do not create or experiment with such places, and therefore, these research subjects have been limited to physical existences in the real world.

In contrast, research studies on human bodies have focused on virtual worlds. 
For example, researchers in the interdisciplinary field of engineering and cognitive science examined the relationship between human bodies and their virtual avatars, such as the Alter Ego\cite{NagakamiTakeuchi2008}.
Their research method, the constructive and experimental approach, aims for understanding through making. This study applies this approach to study the virtual place.

\subsection{Exploratory experiment}


To our knowledge, studies focusing on the equivalence of virtual and real-world places are lacking. However, research studies on déjà vu are instructive in this regard. Déjà vu, which is the odd feeling of having experienced a situation before despite the situation being new\cite{Cleary2009}, has some commonalities with the subject of this study. Both includes some kinds of feeling of equivalence.

In studies that explore factors contributing to déjà vu, experimental participants are often shown static images of scenes and asked to rate some measures related to déjà vu, e.g., familiarity \cite{Cleary2009}.
Some relatively recent studies use 3DCG virtual environments\cite{Cleary2012} as the experimental stimulus.
Such methods are effective for testing the hypotheses of specific known factors; one of the objectives of this study is to present a hypotheses to be tested.


Unknown factors can be identified using different methods. One method involves allowing the experimental participants to act freely and observing their actions. In the photo projective method (PPM), participants are provided a camera and instructed to take photos. The resulting images are considered a reflection of the individual's relationship with the external environment and used to understand the perceived environment and the individual's psychological world\cite{Cherem1983}. Stedman et al.\cite{Stedman2004} attempted to analyze local elements that foster place attachment using this method.
Although this technique is effective, it has the disadvantage in that it reveals only what is seen in the photos.


In another study on places, the participants are asked to place objects freely in the given room. Meagher\cite{Meagher2014} conducted an experiment in which a small office was decorated with stationery and other objects.
Hirose et al. captured a video of the scenery of the experiment, which is basically the same procedure\cite{Hirose2023}.
If this method is used to observe behaviors, unlike PPM, all behaviors can be captured.


Inspired by these studies, we plan to conduct an experiment on how objects are placed freely within a virtual environment because we expect to extract unknown factors from the observation of such free behavior and because the subject of this study is a virtual environment.
Further, in a virtual environment, data logging allows all behaviors to be captured without the need for video recording.

\section{Place Sameness}


The concept of place sameness is proposed in this paper. Place sameness can be defined as the degree of similarity between an original and a virtual re-creation that represents the original.

In this context, the meaning of ``same’’ is not only the visual resemblance or material equivalence to another place, but also its ontological equivalence as an entity.
A place could have a place sameness with another place even if it is materially different, or it is made of 3DCG.


This concept should not be confused with place identity, which is a Developmental Psychology and Environmental Psychology concept, and it deals with the part of a person’s identity associated to places\cite{PlaceIdentity}.
Place sameness is the property of a place. However, because place has an element of subjective meaning, there is a profound interrelation between place identity and place sameness\cite{Relph1976}.

Yet another issue is measurement method used for this concept.
Given the conceptual similarity to déjà vu, familiarity is a measure of place sameness between a place we already know and its virtual re-creation.
Moreover, this concept addresses more than visual similarity; however, because space and material are also components of place, visual similarity is included as an indicator of place sameness.
In addition to this perceptual similarity, cognitive judgments can also be considered an indicator. These judgements decide whether one place and another can be considered the same. These questions are intended to be asked in a questionnaire or in an interview. However, it may eventually require brain function measurements related to spatial cognition.

\section{Exploratory Experiment}



Aoyagi and Fukumori\cite{Aoyagi2022} used 3D static images of a university classroom to examine the semantic proximity between objects and lectures as a factor of place sameness. 
Participants who knew the original classroom were shown an accurate recreated image of a certain classroom, and an image without details, image without student desks, and image without whiteboard, in a random order. 
The participants then answered questions related to place sameness, including ``where is the original place?'' and place sameness indicator items such as the level of confidence in their answer.
The hypothesis was that the image without whiteboard, which is deeply related to lectures as a component of the place, would have the lowest place sameness indicator. 
However, the results showed that the image without desks had the lowest, and the proposed hypothesis was not supported.

Further, the result showed that the image without details had a higher level of confidence in the answer than the accurate recreated image. 
This finding is similar to the uncanny valley effect of humanoid robots, which is the feeling of uncanny when human-like robots closely mimic human but are not quite perfect\cite{Mori2007}.
In addition, this suggests that there is also an uncanny valley effect in place, and it may be considered for achieving place sameness.

Since the above hypothesis was not supported, another hypothesis about factors of place sameness needs to be explored. Participants could easily assume the answer of ``where is the original place?'' because all images shown in the experiment originated in the same room. Therefore, it is possible that the confidence was not measured accurately.
Thus, there is a need for a different experimental design besides showing similar images multiple times.

To this end, an exploratory experiment with a classroom re-creation task in a virtual environment was conducted.
The objectives of the experiment include (1) exploring factors that enhance place sameness by focusing on objects selected by participants and their placement, and (2) revisiting the uncanny valley effect of places using different experimental methods.

\subsection{Method}
This experiment was reviewed by the Komazawa University Ethics Committee on Research Involving Human Subjects.

\subsubsection{Participants}
15 Japanese university students (4 females and 11 males) participated in the experiment. The participants were compensated with 1,000 yen after completing the experiment.

\subsubsection{Materials and Equipment}
The experiment was conducted in a laboratory on the campus of a university to which the participants belong.
The participants accessed the experimental system and operated it using a web browser on a computer.
The computer display was 27 inches with a resolution of 3840 x 2160 pixels.
The experimental system was a combination of virtual environments developed using the Unity game engine\footnote{https://unity.com} and questionnaires generated using JsPsych\cite{jsPsych}.


The experimental system contained one virtual environment for practice and two virtual environments that mimic a real university classroom.
The participants can freely move their viewpoints and place objects in each virtual environment using a mouse.
Information about the original university classrooms and their virtual re-creations is presented in Table \ref{roomInfo}.

\begin{figure}[tb]
  \begin{minipage}{0.5\hsize}
    \begin{center}
     \includegraphics[width=\columnwidth]{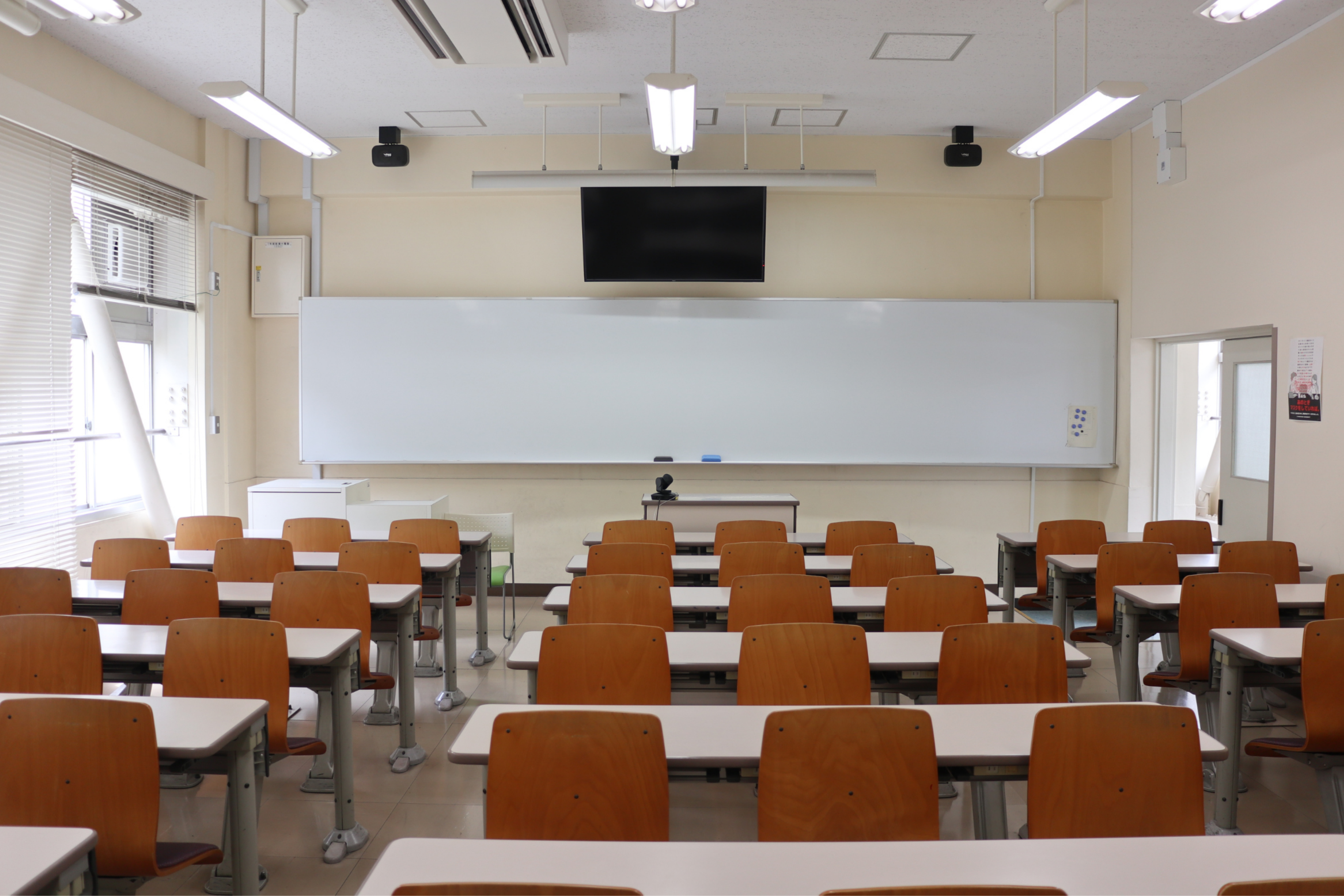}
       \par
    (a) Room A.
    \end{center}
  
  \end{minipage}
  \begin{minipage}{0.5\hsize}
    \begin{center}
     \includegraphics[width=\columnwidth]{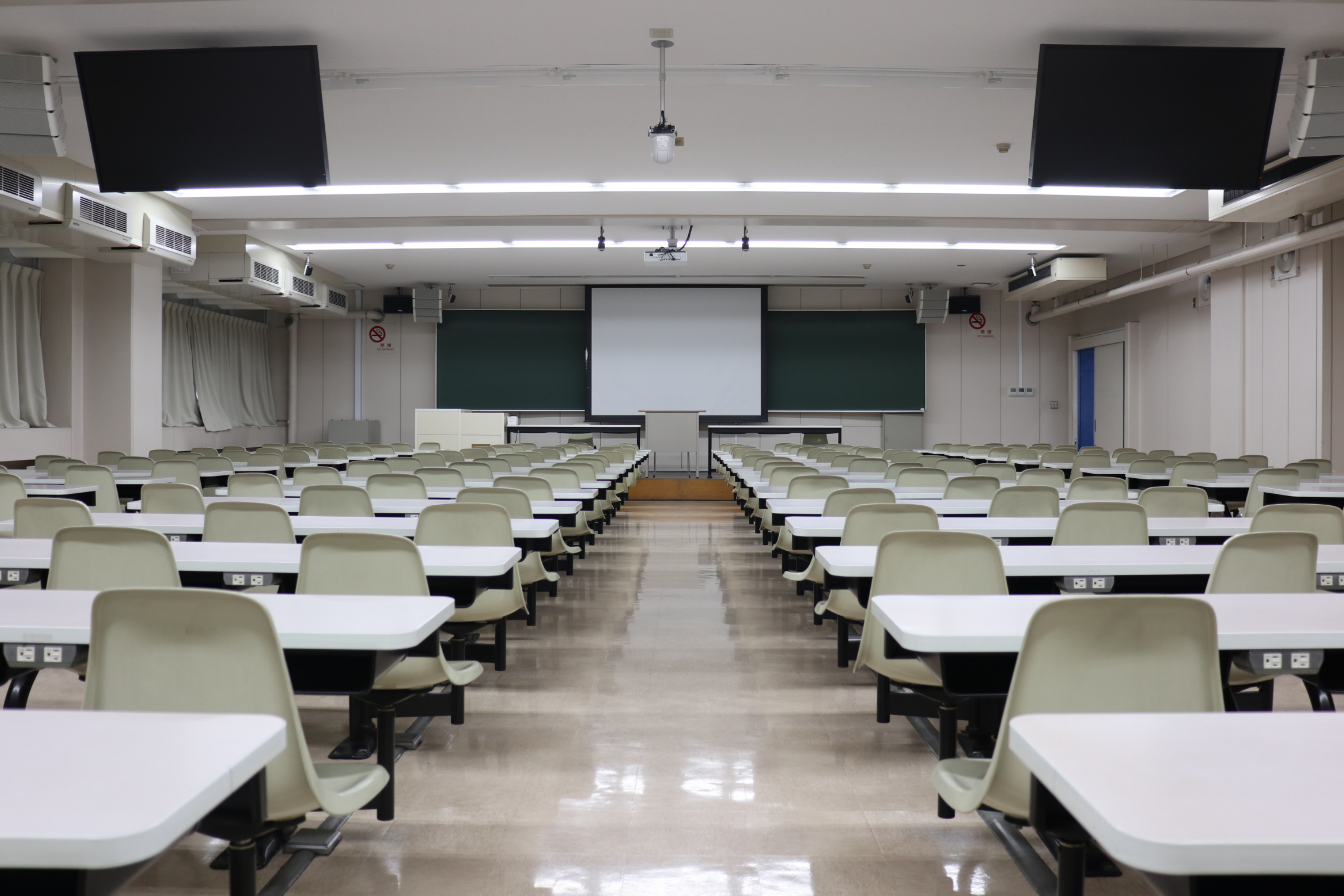}
       \par
    (b) Room B.
    \end{center}
   
  \end{minipage}
  \caption{Pictures of rooms A and B.}
  \label{pictures}
\end{figure}

\begin{figure}[tb]
  \begin{minipage}{0.5\hsize}
    \begin{center}
     \includegraphics[width=\columnwidth]{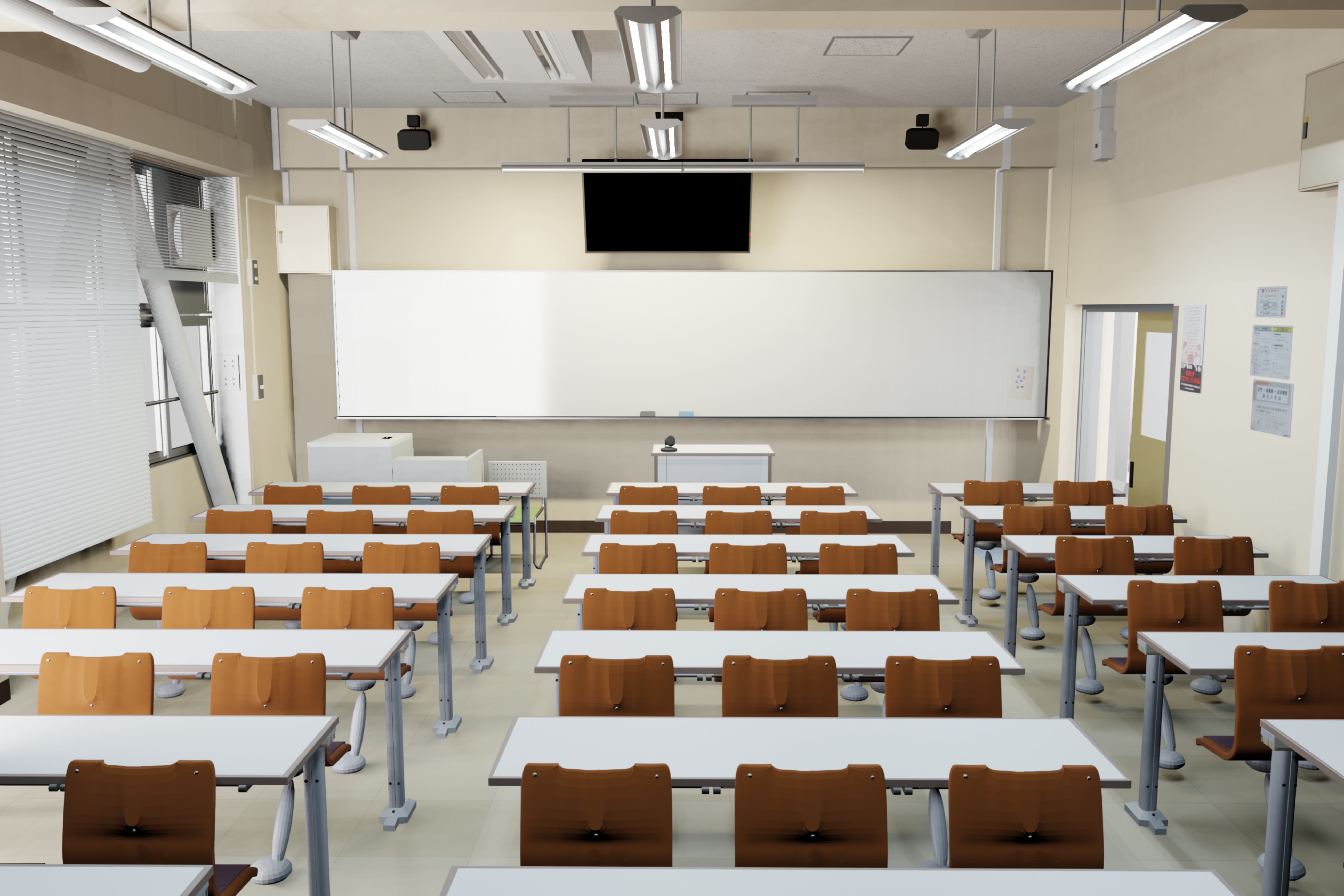}
      \par
    (a) Room A.
     \end{center}
    
  \end{minipage}
  \begin{minipage}{0.5\hsize}
    \begin{center}
     \includegraphics[width=\columnwidth]{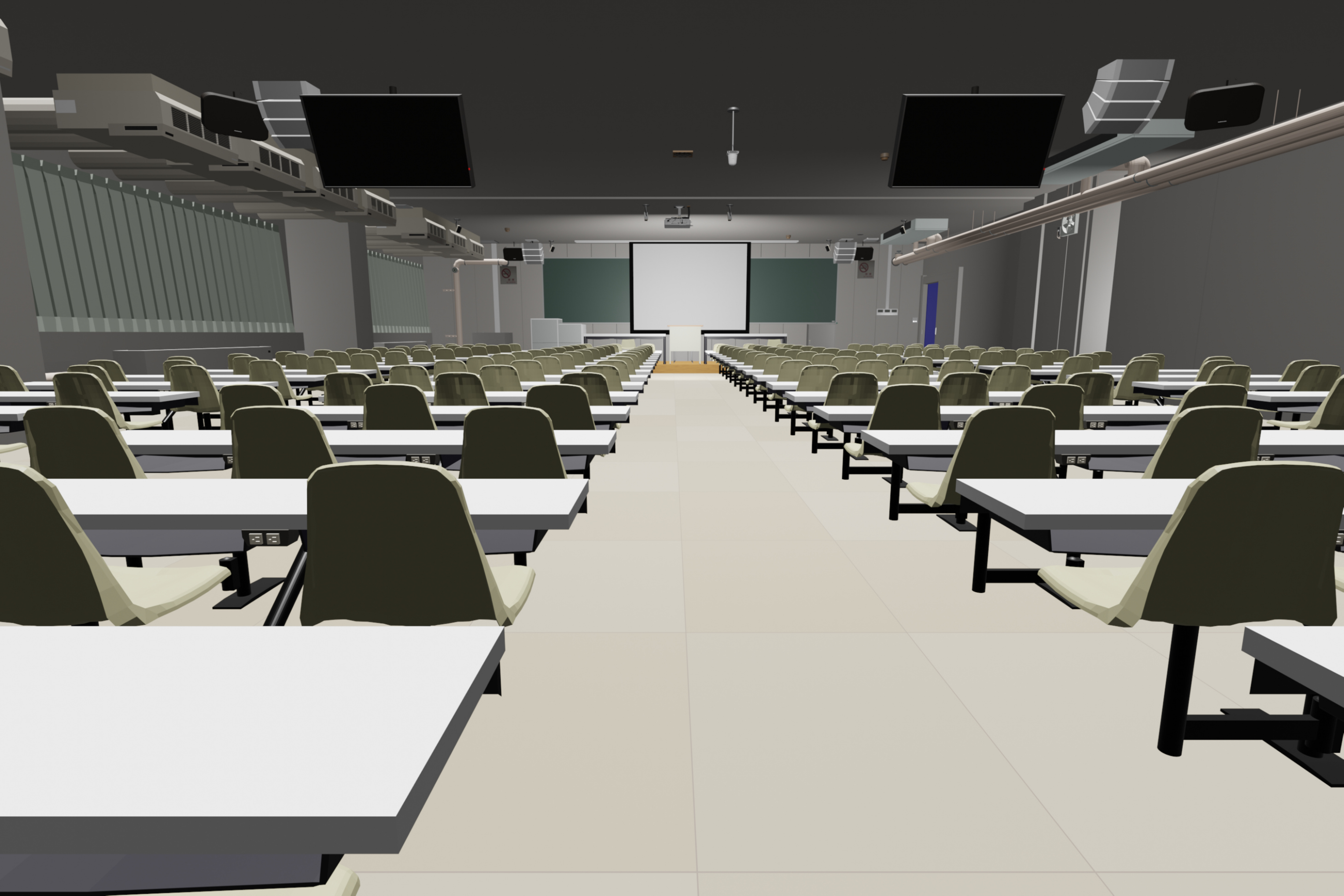}
     \par
    (b) Room B.
    \end{center}
    
  \end{minipage}

  \caption{Base models of rooms A and B.}
    \label{models}
\end{figure}

Fig. \ref{pictures} (a) shows a picture of Room A, and Fig. \ref{pictures} (b) shows a 3DCG model of Room A.
Room A is a small classroom that is basically the same as that in Aoyagi and Fukumori’s study \cite{Aoyagi2022}. This room is used for small lectures and seminars.
Fig. \ref{models} (a) shows a picture of Room B, and Fig. \ref{models} (b) shows a 3DCG model of Room B. Room B is a large classroom and is used for large group lectures.

The number of objects differs from the number of types of objects because some of the room fixtures had more than one installation.
The object types were divided into three categories. ``Display devices'' included whiteboards, blackboards, and displays; ``Student desks'' included desks and chairs for students; and ``Others'' included all remaining objects such as switch boxes, cameras, outlets, or fluorescent lamps. 
To consider activities as the components of a place, the display devices are related to the lecture and the seminar; the student desks, seminars, chatting, and self-study; and others, not related to any activities. 

\begin{figure}[tb]
  \begin{minipage}{0.5\hsize}
    \begin{center}
     \includegraphics[width=\columnwidth]{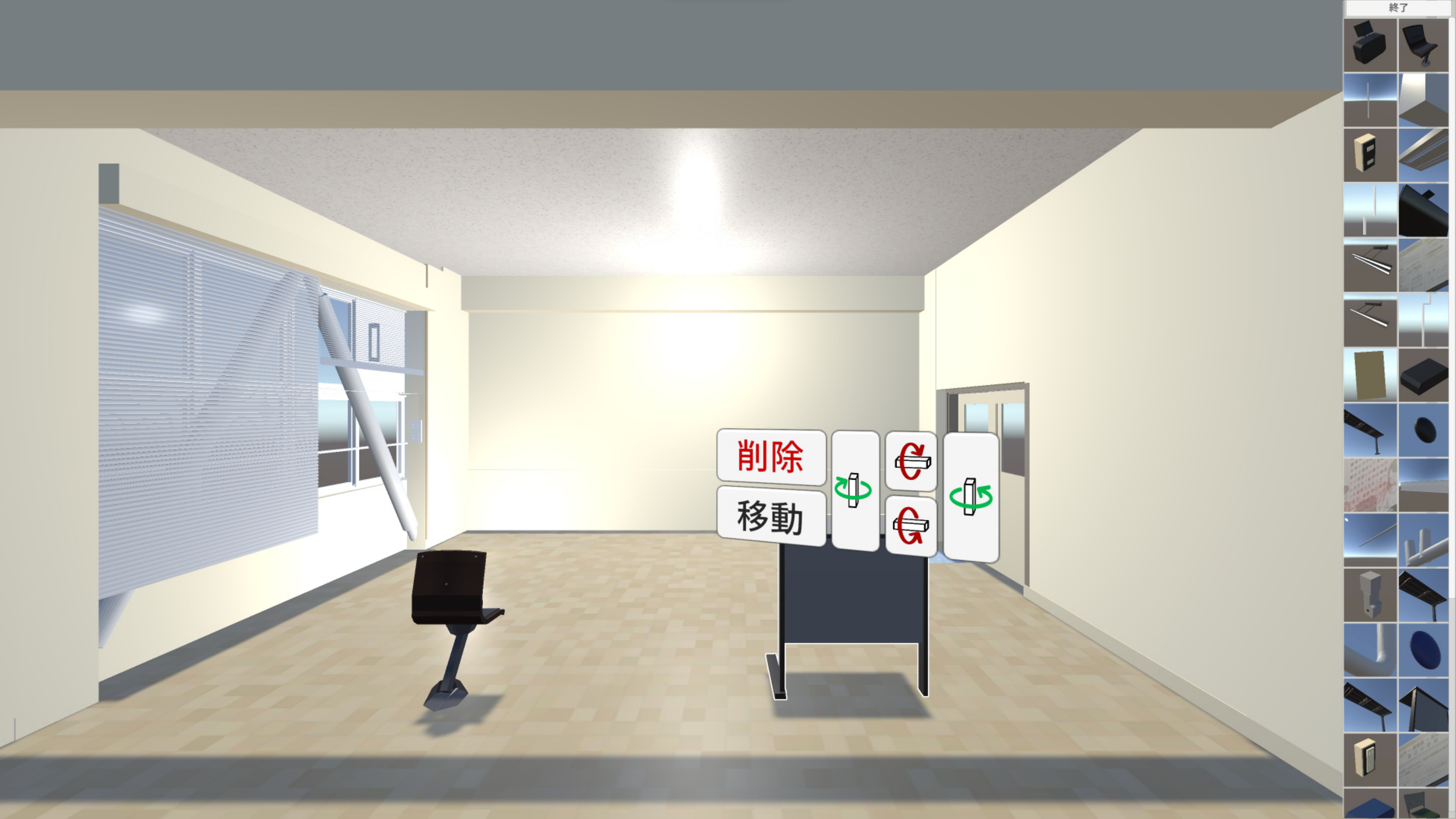}
       \par
    (a) Room A.
     \end{center}
  \end{minipage}
  \begin{minipage}{0.5\hsize}
    \begin{center}
     \includegraphics[width=\columnwidth]{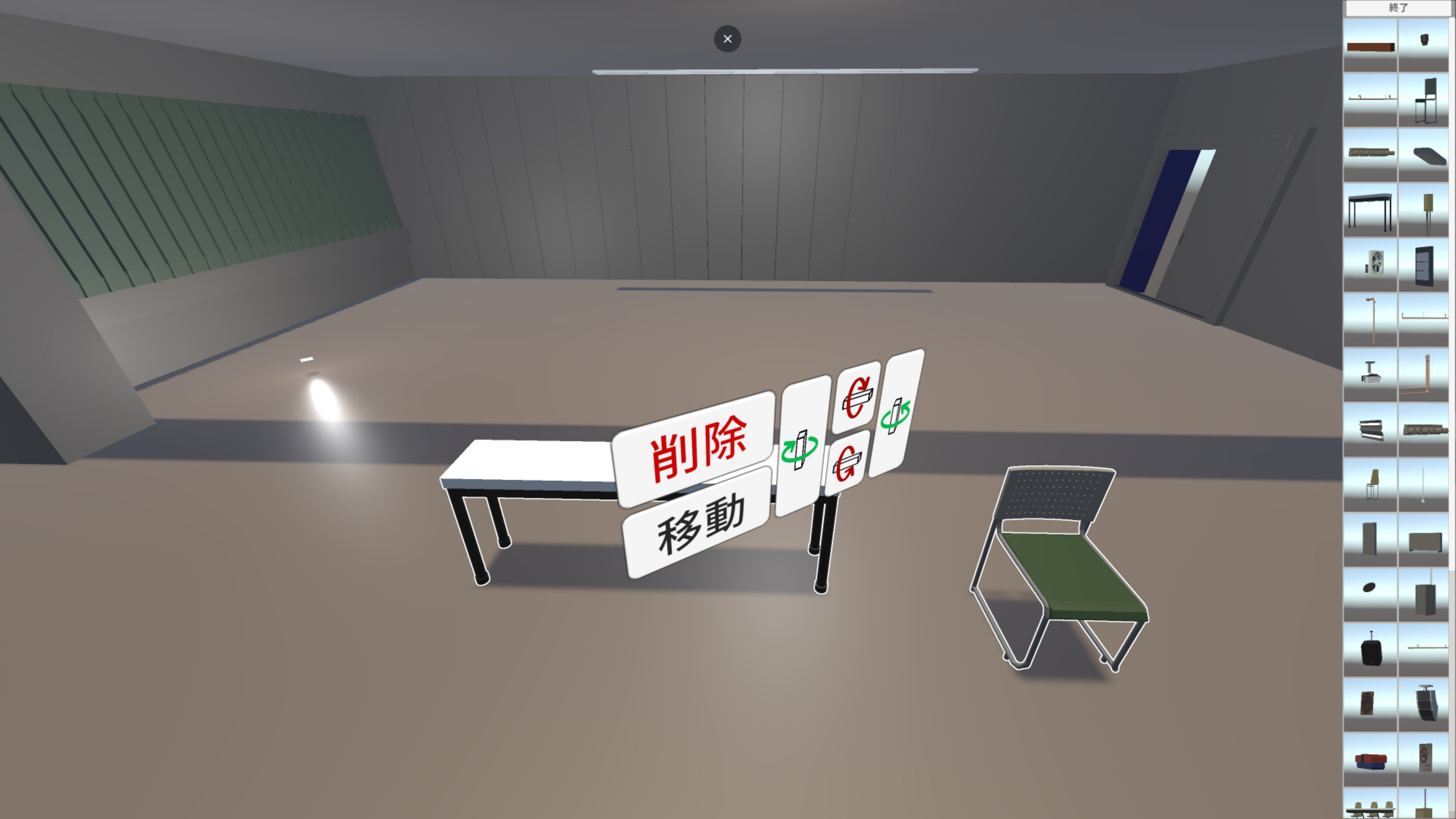}
       \par
    (b) Room B.
    \end{center}  
  \end{minipage}
  
  \caption{Experimental scenes.}
    \label{scenes}
\end{figure}

In the classroom re-creation task in virtual environments, all objects except walls, floors, and ceilings were removed from the 3DCG models of the rooms in advance.
Figs. \ref{scenes} show experimental scenes of Rooms A and B. Fig. \ref{practice_scene} shows a virtual environment for practice. The participants can freely place, move, and rotate objects by clicking object icons on the right side on the screen.
This system enables the participants to recreate the real rooms in the virtual environments.

\begin{table}[tb]
\caption{Specifications of rooms}
\begin{center}
\label{roomInfo}
\begin{tabular}{lrr}
\hline
                           & Room A             & Room B               \\
                           \hline
Number of seats & 60   & 272   \\
Height[m]       & 3.66 & 3.34  \\
Width[m]        & 6.72 & 13.87 \\
Depth[m]      & 9.1  & 17.72\\ 
\hline \hline
Number of object types & 39 & 42  \\
\hline
Display devices [\%]  & 7\% & 5\% \\
Student desks [\%] & 2\% & 3\%\\
Others [\%] & 97\%  & 90\% \\
\hline \hline
Number of objects  & 105 & 151  \\
\hline
Display devices [\%]  & 3\% & 2\% \\
Student desks [\%] & 43\% & 45\%\\
Others [\%] & 55\%  & 52\% \\
\hline
\end{tabular}
\end{center}
\end{table}

\subsubsection{Procedure}



All instructions and questionnaires were conducted in Japanese. Informed consent was obtained from the participants before the experiment.

The participants sat in a chair in front of a desk and faced the computer display on the desk. First, the participants viewed a screen which describes the procedure of the experiment and provides instructions on how to operate the experiment system. Meanwhile, the experimenter also explained the same content. This screen was also displayed on a tablet device placed on the desk, and it was possible to always look back.

\begin{figure}[tb]
\begin{center}
\includegraphics[width=0.6\columnwidth]{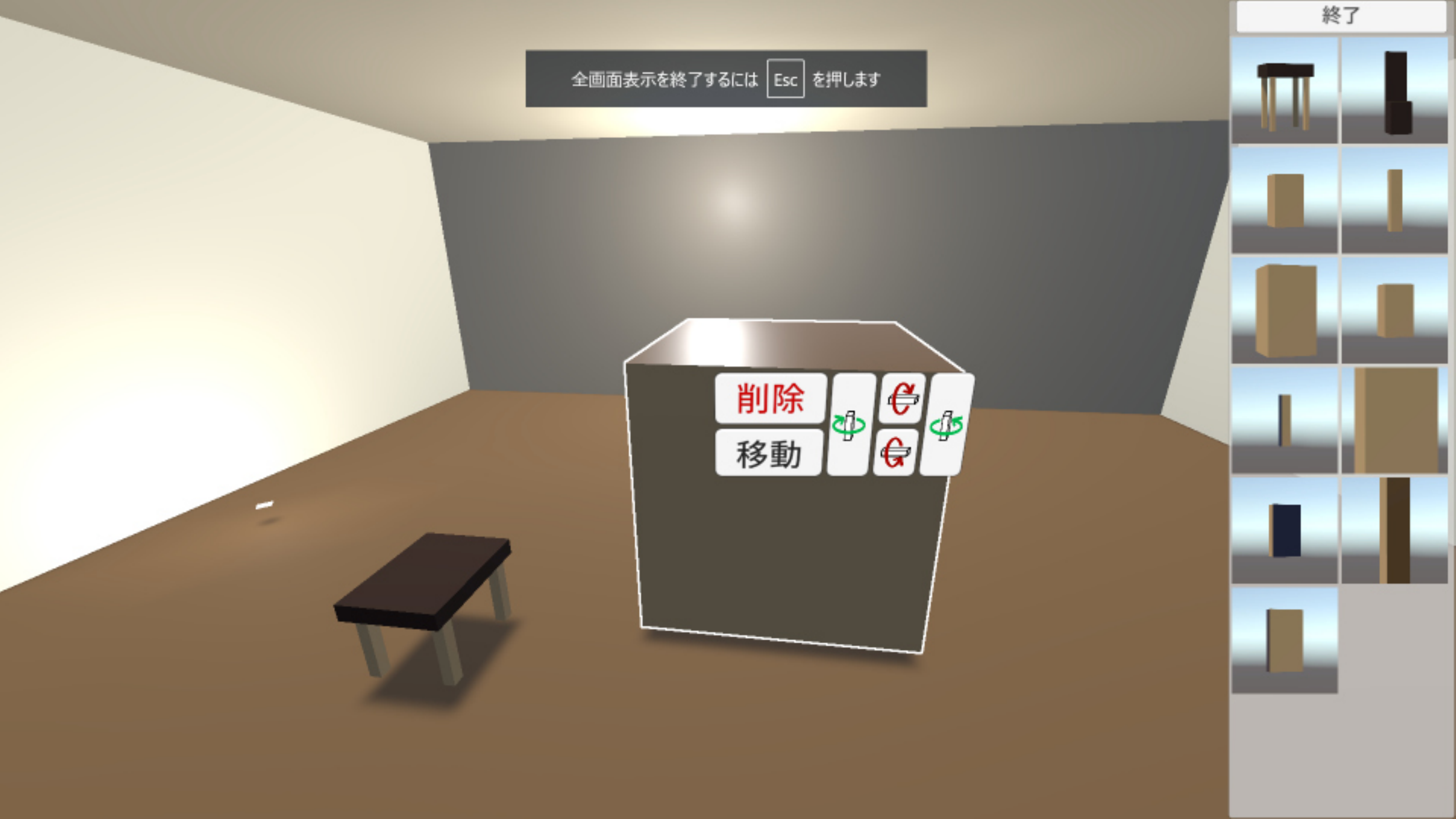}
\end{center}
\caption{Virtual environment for practice.}
\label{practice_scene}
\end{figure}

On the next screen, the participants were presented with a virtual environment for practice (Fig. \ref{practice_scene}).
In all virtual environments, including this one, the participants can place objects shown in the right area on the screen (Fig. \ref{practice_scene}).
The participants can proceed to the next screen at any time by pressing the end button on the screen.
10 min after starting the screen, it automatically advances to the next screen.
If 10 min have passed since the screen started and it has not yet advanced to the next screen, it is automatically advanced to the next screen.

Next, the participants engaged in the classroom re-creation task in virtual environments of Rooms A (Fig. \ref{scenes} (a)) and B (Fig. \ref{scenes} (b)) in random order.
The participants were instructed to mimic the original room using objects shown in the right pane on the screen (Fig. \ref{scenes}).
All movements and rotations of the viewpoint and objects were logged with timestamps.

After each tasks was finished, the participants answered the questionnaire.
The questionnaire included questions such as ``how attached are you to that room?'',``how often did you use that room?,'' and ``when was the last time you used that room?'' about the original rooms and rooms similar to the original rooms listed in Tables \ref{items} and \ref{memorable}.
This is because the rooms within each building of this university are very similar to each other.
If a participant did not know the original room, and instead knew a similar room, they could re-create the room quite accurately.
If a participant did not know both of them, the results would be based on creation, and not re-creation.
Therefore, it is necessary to confirm whether the participants knew original or similar rooms.
In addition, items of the place sameness index were answered in random order.
The place sameness index includes seven items, which are as follows: It looks like the original place; It is similar to the original place; I can tell it is a re-creation of the original place; It looks familiar; I feel uncomfortable (a reversing entry); I successfully recreated it; and I am satisfied with the result.
The participants answers these questions using a 5-point scale: strongly disagree, disagree, neutral, agree, and strongly agree.

\subsection{Results}

Figure \ref{items} shows a summary of the answers to ``how attached are you to that room?'', ``how often did you use that room?'', and ``when was the last time you used that room?''.
Attachments to the original and similar rooms were not considerably different. Room A had a higher attachment than that of Room B.
The most common answer for frequency was once a week for all rooms;
the most common answer for ``when was the last time you used that room?''  was ``1 month to 1 year ago.''
On the other hand, that of Room B was ``today or yesterday.''

{
\begin{table}[tb]
\caption{Usage of each room in the experiment}
\begin{center}
\label{items}
\tabcolsep = 3pt
\begin{tabular}{p{2.3cm}p{1.1cm}p{1.1cm}p{1.1cm}p{1.1cm}}

\hline
                & \multicolumn{2}{c}{Room A} & \multicolumn{2}{c}{Room B} \\
                & Original
                &  Same type
                & Original
                &  Same type  \\
                \hline
                \\[-0.8em]
How attached are you to that room? \small{*1}       &  2.4(1.32)           & 2.13(0.71)        & 1.67(1.05)     & 1.67(0.97)     \\
\\[-0.8em]
How often did you use that room? \small{*2*3}              &once a wk   (12)    & once a wk (11)    & once a wk (11) & once a wk (12) \\
\\[-0.8em]
\multicolumn{1}{p{2.3cm}}{When was   the last time you used that room? \small{*2*4}  } & 1 mth to 1   yr (7) & 1 mth to 1 yr (5) & ytdy (6)       & ytdy (8)          \\
\hline
\multicolumn{5}{p{8cm}}{*1: Average of 5 points scale: not at all, not very much, neutral, moderately, and very much (standard deviation)}   \\
\multicolumn{5}{l}{*2: Most frequent answer (number of answers)}    \\
\multicolumn{5}{l}{*3: Once a wk: about once a week }        \\
\multicolumn{5}{l}{*4: Ytdy:  today or yesterday, 1 mth to 1 yr: 1 month to 1 year ago} 
\end{tabular}
\end{center}
\end{table}
}

Table \ref{memorable} shows answers to ``what is the most memorable activity in the room?''
The most common answer was ``lecture'' for all rooms.
Answers for Room A were more diverse than those for Room B. Three participants answered ``seminar'' for original Room A and similar rooms.
However, for Room B, only one participants answered ``seminar'' for similar rooms.

\begin{table}[tb]
\caption{Answers to ``what is the most memorable activity in the room?''}
\begin{center}
\label{memorable}
\begin{tabular}{lp{1cm}p{1cm}p{1cm}p{1cm}}
\hline
                & \multicolumn{2}{c}{Room A} & \multicolumn{2}{c}{Room B} \\
                & Original
                &  Similar
                & Original
                &  Similar  \\
                \hline
Club       & 0 & 0 & 0  & 0  \\
Seminar    & 3 & 3 & 0  & 1  \\
Lecture    & 9 & 9 & 14 & 13 \\
Chatting   & 1 & 1 & 0  & 0  \\
Self-study & 0 & 0 & 0  & 0  \\
Others     & 0 & 0 & 0  & 1  \\
Never used & 2 & 2 & 1  & 0  \\
\hline
\end{tabular}
\end{center}
\end{table}

Place sameness index includes seven items, and the value for Cronbach's coefficient alpha $ \alpha = 0.86 $. The averages of the seven items are used as the values of the place sameness index in the following analysis.

To explore the factors of place sameness index, some variables were calculated based on logging data. Some participants placed too many objects in the virtual rooms, and others placed too few, compared to the original rooms.
This difference in behavior could be attributed to memory.
In this study, the value calculated as $ |1-\frac{N_{p}}{N_{c}}|,$ where $N_{p}$ and $N_{c}$ represent the number of objects of re-creations and the number of objects in each original room, respectively, is used as a measure of how well the participants remembered the objects.
This value represents how accurately the number of real-world objects is remembered, and hereinafter, it is referred to as the memory index.
A memory index of 0 implies that the number of objects in the virtual room perfectly match with that in the original.

The positions of placed objects in the virtual rooms were not always accurate. The average distances were calculated to represent the accuracy of re-creations in terms of the positions of objects.
It is defined as the per-participant averages of Euclidean distances between the positions of all objects at the end of the experiment and positions of the objects in each original room. 

The third variable is the average placement order of objects.
Recalling important objects is believed to be easy, and they are often placed early. These three variables were calculated for different object types (display devices, student desks, and others).

\begin{figure*}[tb]
\begin{center}
\includegraphics[width=\textwidth]{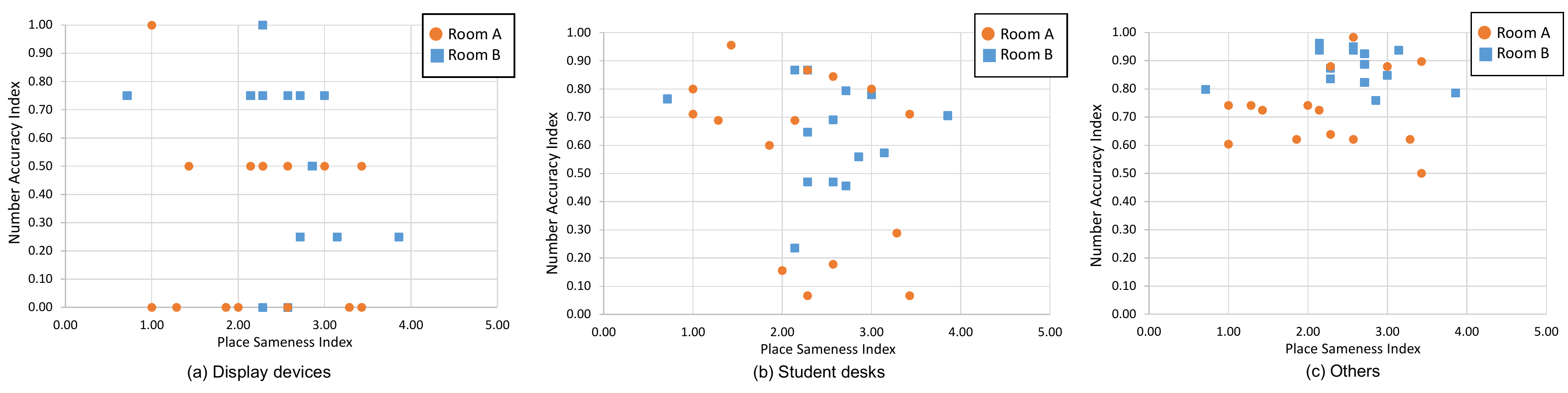}
\end{center}
\caption{Memory index and place sameness index.}
\label{propotion_of_objects}
\end{figure*}

\begin{figure*}[tb]
\begin{center}
\includegraphics[width=\textwidth]{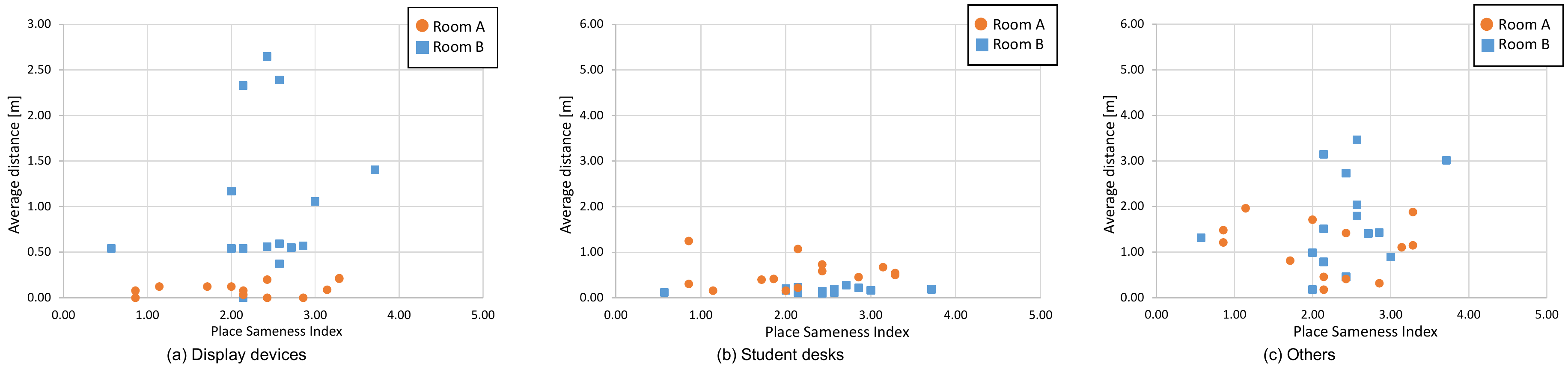}
\end{center}
\caption{Average distances of object categories and place sameness index.}
\label{average_distance}
\end{figure*}

\begin{figure*}[tb]
\begin{center}
\includegraphics[width=\textwidth]{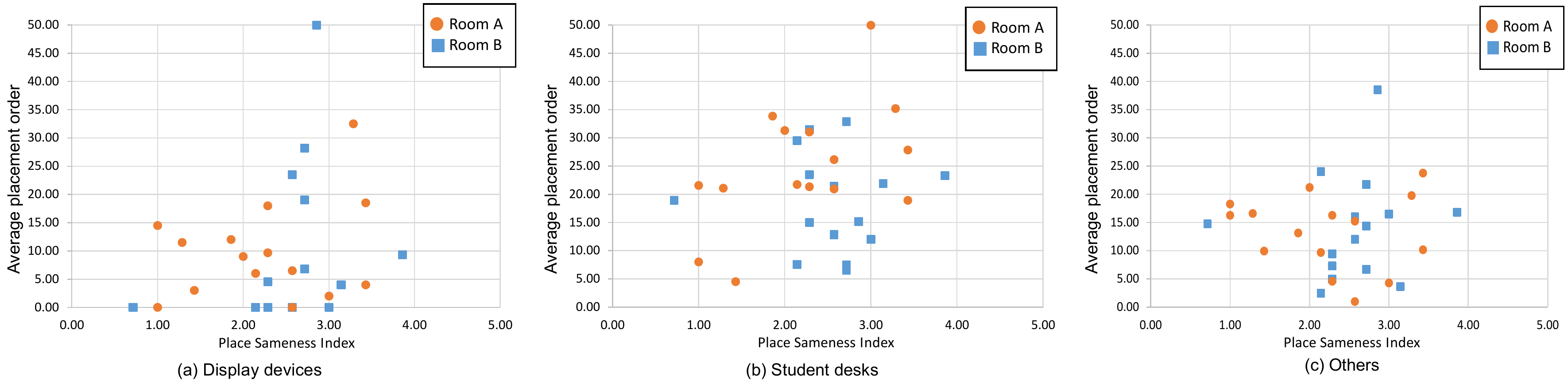}
\end{center}
\caption{Average placement orders and place sameness index.}
\label{average_placement_order}
\end{figure*}

Figures\ref{propotion_of_objects},  \ref{average_distance}, and \ref{average_placement_order} present the scatter plots of place sameness index and memory index, average distance, and average placement order for each room and each category of object types.
The variables for each room and each category showed different trends.
For example, the average distance of others of Room A shows a downward trend, and that of Room B shows an upward trend.
To make this difference easier to understand, a correlation analysis was conducted between each place and the place sameness index.

Table \ref{correlation} shows the correlation coefficients of the variables and the place sameness index.
To calculate correlation, the values for one participant who reversed the front and back of Room A were excluded as outliers. Criteria defined by Mizumoto and Takeuchi \cite{EffectSize1} based on \cite{EffectSize2} is used to interpret effect sizes.
In the criteria, 0.1, 0.3, and 0.5 represent small, medium, and large effect sizes, respectively. The values with medium and large effect sizes are presented in bold in Table \ref{correlation}.

Moderate or large correlations were observed for 
The display devices in Room B, the student desks in Room A at the memory index, display devices in Room A, and student desks and others in Room B at the average distance, and student desks in Room A at the average placement order.

\begin{table}[tb]
\caption{Correlation coefficients of variables and place sameness index}
\begin{center}
\label{correlation}
\begin{tabular}{p{1.1cm}cccc}
\hline
&  &\multicolumn{3}{c}{Correlation coefficient r *1} \\
  Variables  & Category     & Whole data     & Room A         & Room B        \\
                      \hline
\multirow{3}{1.1cm}{Memory index}      & Display devices          & -0.16          &- 0.14          & \textbf{-0.38} \\
                                           & Student   desks & -0.23  & \textbf{-0.39}  & -0.04  \\
                                           & Others          &  0.16 & 0.10 & -0.05 \\ 
                                             \hline
\multirow{3}{1.1cm}{Average   distance}        & Display devices & 0.19           & \textbf{0.53}  & 0.14  \\
                                           & Student   desks          & -0.07          & 0.01           & \textbf{0.33}  \\
                                           & Others                   & 0.15           & -0.20          & \textbf{0.30}  \\
                                           \hline
\multirow{3}{1.1cm}{Average placement order} & Display devices          & 0.21           & 0.16           & 0.27  \\
                                           & Student   desks& 0.22           & \textbf{0.52}  & -0.02 \\
                                           & Others  & 0.02 & -0.09 &0.08  \\\\ 
                                            \hline
\multicolumn{5}{l}{*1: Values with medium and large effect sizes are bolded.} \\
\end{tabular}
\end{center}
\end{table}

Table \ref{objects_order} shows the average placement orders of objects and the percentages of the number of participants who placed the object to the number of all participants.
The table includes objects placed by more than half of the participants. The objects are arranged in the ascending order of average placement order.
Ten types of objects of Room A and 3 types of that of Room B are shown in Table \ref{objects_order}.
A comparison of display devices and student desks indicates the tendency for the former to be placed relatively early. In addition, Room A has more others objects than those in Room B.

\begin{table*}[tb]
\caption{Average placement orders and percentages of the number of participants who placed the object}
\begin{center}
\label{objects_order}
\begin{tabular}{llp{1cm}p{1.1cm}|llp{1cm}p{1.1cm}}
\hline
\multicolumn{4}{c|}{Room A}                                                              & \multicolumn{4}{c}{Room B}   \\
\hline
Object type & Category & Average order & Rate of participants  & Object type & Category   & Average order & Rate of participants \\
\hline
Teacher's desk & Others          & 8.00  & 100\%                         & Blackboard          & Display devices & 11.13 & 53\%                          \\
Camera                & Others          & 10.53 & 80\%                          & Side teacher's desk & Others          & 16.50 & 53\%                          \\
Switch box            & Others          & 11.64 & 67\%                          & Student's desk      & Student desks   & 21.91 & 100\%                         \\
Whiteboard            & Display devices & 11.92 & 73\%                          &                     &                 &       &                               \\
Long student's desk   & Student desks   & 18.30 & 60\%                          &                     &                 &       &                               \\
Speaker               & Others          & 21.18 & 73\%                          &                     &                 &       &                               \\
Short student's desk  & Student desks   & 21.87 & 80\%                          &                     &                 &       &                               \\
Ceiling lighting      & Others          & 23.50 & 60\%                          &                     &                 &       &                               \\
Wall objects A        & Others          & 28.25 & 53\%                          &                     &                 &       &                               \\
Student's chair       & Student desks   & 38.89 & 93\%    \\
\hline

\end{tabular}
\end{center}
\end{table*}

\subsection{Discussion}

The original classrooms were known to most experimental participants;
they had used them about once a week (Table \ref{items}).
In addition, they used the original or similar rooms in the recent past.
Thus, it can be inferred that most participants remembered and were able to re-create the rooms.


Room B was used only for lectures, and Room A was used for lectures, seminars, and chatting.
Answers for Room A were slightly diverse.
This difference in the nature of the classrooms indicates that activities as components of places are different (Fig. \ref{memorable}).
The results of the correlation analysis can be interpreted in this light (Table \ref{correlation}).


For the memory index, student desks in Room A (Fig.\ref{propotion_of_objects} (b)) show a moderate negative correlation with place sameness index. That of Room B does not show that. 
During any activity in the classrooms, the object closest to the students is their desk and chair.
However, the accurate re-creation of the number of desks and chairs in Room A was correlated to the place sameness index, but not in Room B.
This may be because students are more impressed with the desks and the chairs in Room A and less so in Room B.

Based on these results, the following inferences can be drawn.
In activities reported only in Room A, i.e., seminar and chatting, students saw other students and the desks and the chairs they were sitting in, and therefore, the desks and the chairs were impressive.
Room B was used mostly for lectures, and therefore, the students did not watch other students' desks and chairs, and instead focused on the teacher and the whiteboard. Therefore, the chairs were not impressive.
In other words, the correct number of objects related to the activity of the room has a positive effect on place sameness.

The same inference can be drawn for display devices.
Since Room B is used for lectures, students focused on the whiteboard, and therefore, the re-creation of that number has a moderate negative correlation with the place sameness index.
Room A is not used only for lectures, and therefore, the correlation of the same value is small in comparison.


The existence of the uncanny valley effect of place was supported by a weak correlation of others in Room A, which is the same as that reported by Aoyagi and Fukumori\cite{Aoyagi2022}.
The original of this effect refers to a nonlinear curve in the accuracy of a re-creation and the sense of uncanniness.
However, in this study, we wanted to interpret it in terms of whether a re-creation appears accurate.
Therefore, the uncanny valley effect can be paraphrased as ``sometimes a more objectively inaccurate re-creation can appear subjectively more accurate than a more objectively accurate re-creation.'' 
Because there are more others objects than the other types, an objectively accurate re-creation of a classroom requires these objects.
However, the more inaccurately the number of objects of others are reproduced, the greater is the place sameness index (Table \ref{correlation}).
This result fits the paraphrased uncanny valley effect.

However, no correlation is found in Room B, and therefore, the uncanny valley effect is not observed. One possible interpretation is that Room B is large, and the participants may not have been able to observe its details.


Some puzzling results were found regarding the distance from the correct answer.
There is a large positive correlation of display devices only in Room A (Fig.\ref{average_distance} (a)), moderate correlations for student desks and others in Room B (Fig.\ref{average_distance} (a)).
The results suggest that re-creating objects with accurate positions is a disincentive for realizing place sameness.

Lectures, seminars, and chatting were held in Room A.
The further away display devices were placed from correct positions (considered to have a smaller relationship with seminar and chatting), the higher was the place sameness.
Only lectures are held in Room B. The further away student desks are placed from the correct positions, which are considered to have little relationship with the Room where only Lectures are held, and from the original location, the higher is the place sameness.
In the opposite direction, object placements are less relevant for placing activities in precise positions may reduce place sameness.



It is unlikely that this trend will hold true on a broader scale.
For example, does placing an object so far away that it cannot be seen enhance place sameness?
This could be part of nonlinear curves like uncanny valley. This should be verified in larger places and with many experimental participants.


For the uncanny valley effect, the results were only supported in Room B with respect to the position (Fig.\ref{average_distance} (c)).
In Room B, the more inaccurate the representation of the position of others objects, the higher was the place sameness index.
However, in Room A, the more accurately the position of objects was to others, the higher was the place sameness index.
This can be interpreted from the size of the rooms because Room A is smaller, the participants may have remembered more details, and the required level of accuracy may have been higher.


There is a large positive correlation between the values of the average order of the student desk values and place sameness index only in Room A (Table \ref{correlation}).
In other words, the later the desks were placed, the higher was the place sameness. This is difficult to interpret; however, it may be a consequence of the fact that student desks were paid more attention to in Room A then that in Room B, in the same light as the memory index and average distance.
The higher the number of desks placed, the larger is the average rank.
The results suggests that many desks enhanced place sameness in Room A, whereas many desks did not that in Room B.


Finally, we analyzed the order and categories of objects placed by half the participants (Table \ref{objects_order}).
A comparison of display devices and student desks indicates that there is a tendency for the former to be placed relatively early. Display devices may be relatively more important in the place memory because the recall of the early placed objects may have been fluent. In addition, Room A has a higher number of others objects than that in Room B. Room A is small, and therefore, the other objects may be easier to remember. This can explain why the uncanny valley effect of the place was not observed in Room A. However, the order of placement is influenced by the re-creation task, and therefore, this cannot be determined.


Overall, it is interesting to note the different trends for each variable in Rooms A and B. 
Even if a particular phenomenon such as an uncanny valley effect occurs, its effects can vary depending on the size of the room.

Aoyagi and Fukumori\cite{Aoyagi2022} hypothesized that location identity would decrease as objects closely related to classroom activities, i.e., teaching, were removed, but this hypothesis was rejected.
The study above only dealt with one room and assumed that the primary activity was lecture.
However, this study examined each of the activities that consists each of the two rooms.
By comparing the objects associated with each of the activities and the place sameness index, the results suggest a correlation between place activity and place sameness.

A study in the field of architecture suggests that it is important for the elderly to bring some object of action with them when they move into a residential facility to maintain their quality of life and prevent dementia\cite{Koga2022}.
This example shows the importance of not changing the presence of objects that reflect activity when one's location changes in the real world, which may be interpreted as maintaining place sameness.
The results of this study suggest that this is true not only in real-world places but also in those of the virtual world.


This study has several limitations. 
For example, no statistical validation was conducted because of the small number of participants in the experiment. 
Further, the implications of this study have not been tested.


In addition, participants in this experiment recreated places by themselves, and therefore, they may have liked the re-creations and rated it good on all measurements.
This makes it difficult to separate the judgment of being the same from a good evaluation.
Also, the experiment dealt with the university classrooms and the university students.
The possibility of generalizing the discussion to other types of places is an important issue.

\section{Conclusion}

Place sameness was conceptualized as the first step to addressing equivalence between virtual and real places. The virtual re-creations of university classrooms were created, and an exploratory experiment with student participants were conducted to explore the types of objects and identify where they are placed to reflect the activity and meanings and contribute to achieve place sameness. Finally, some implications about place sameness factors were extracted from the experimental results.




For objects that reflect activities in places, re-creating the correct number of objects or a lot of the objects was suggested to be a factor of place sameness. For object positions, imprecision contributes to the place sameness for objects not related to the activity as a component of the place. 
Our own interpretation of the uncanny valley effect of place was partially supported; however, it was not conclusive. This is possible because only a part of the nonlinear curve of this effect can be observed because of room sizes.
The main contributions of this study are the proposal of the concept of place sameness as a new perspective for virtual re-creation research and the finding of promising factors for that.

In the future research, a further approach can be effective to recreate the social activity by bringing in the virtual characters of the participants or others.
Another possibility is to make a place which has a specific meaning through instructions or tasks, rather than addressing objects.

The difficulty of dealing scientifically with the experience of a place inherently subjective and probably different for each individual is a challenge that still needs to be overcome.
Given that the study on place sameness is in its nascent stage, the authors intend to expand on this work.

\section*{Acknowledgements}
This work was supported by JSPS KAKENHI Grant Number JP20K20121.




\profile[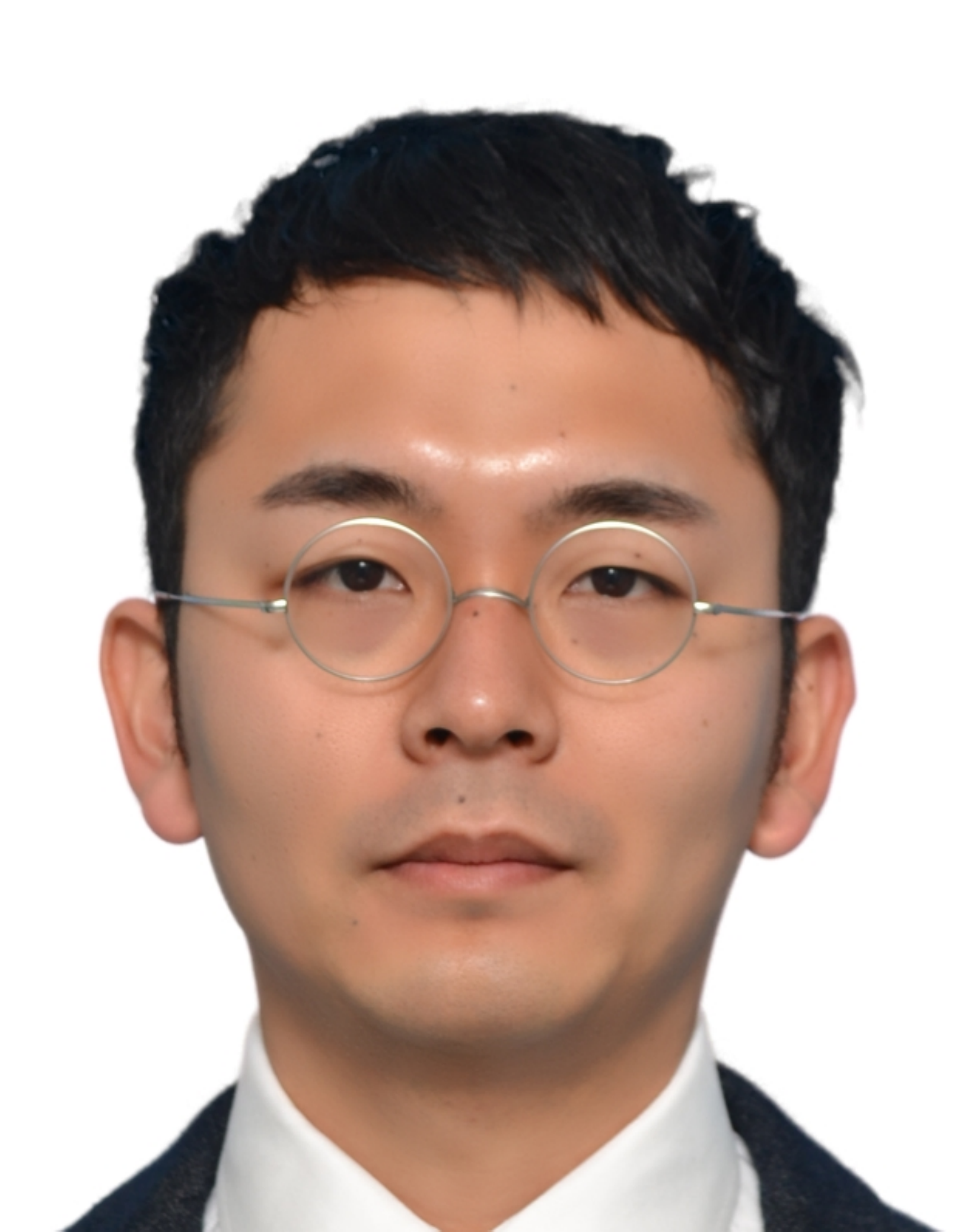]{Saizo Aoyagi}{
received the BE degree in engineering from Kyoto University in 2008.
He received his ME and PhD degrees in energy science from Kyoto University in 2010 and 2012, respectively.
He is presently a lecturer at Komazawa University.
His research interests include the virtual re-creation of places, embodied media, and behavior change using information technology.
He is a member of IEICE, IPSJ, and Human Interface Society.
}

\profile[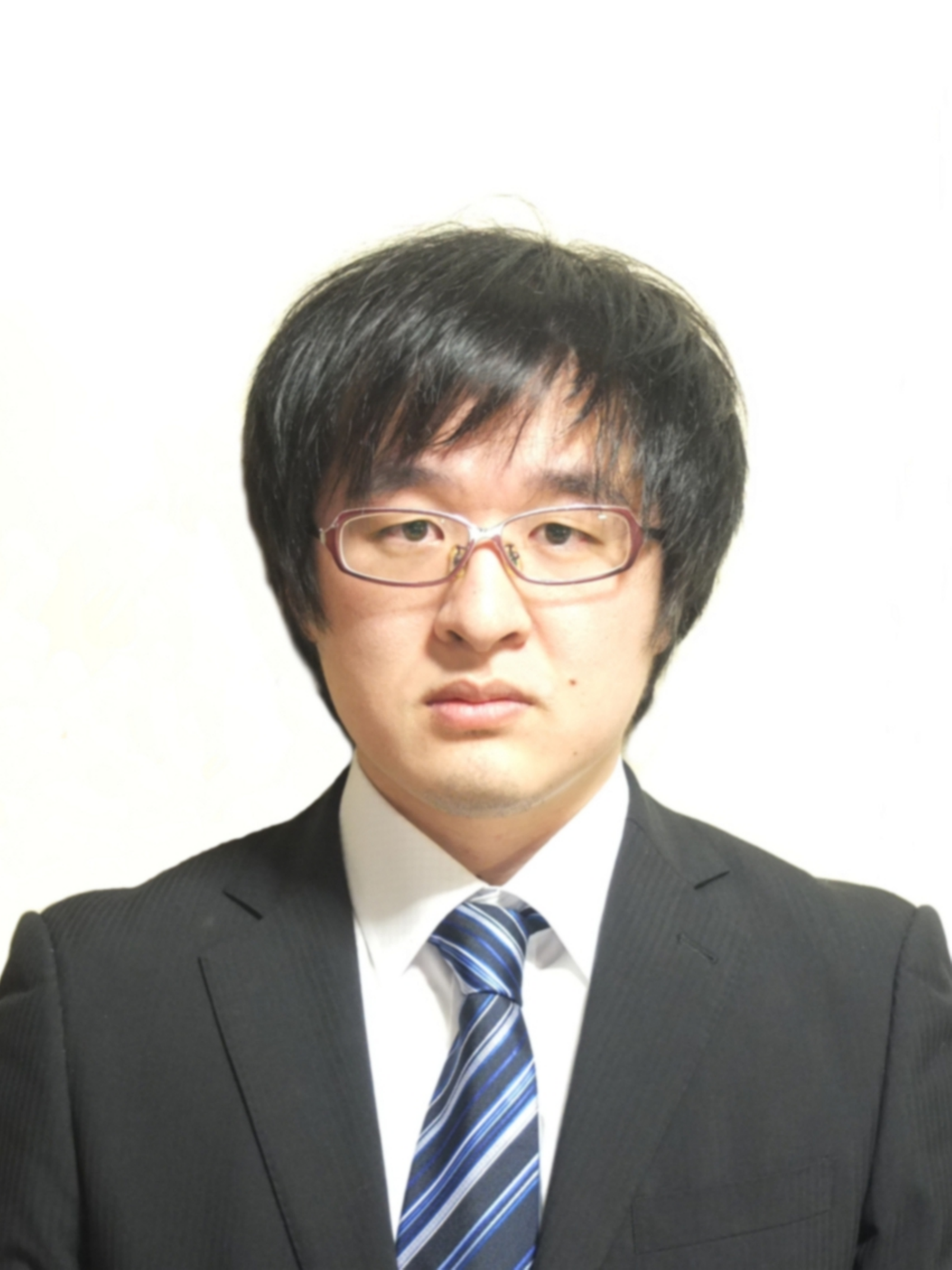]{Satoshi Fukumori}{
received the BE, ME, and PhD degrees in engineering from Okayama University in 2008, 2010, and 2015, respectively. 
He is currently a lecturer at Kagawa University. 
His research interests include human body cognition and its medical applications. 
He is a member of IPSJ, JCSS, and the Human Interface Society.
}

\profile[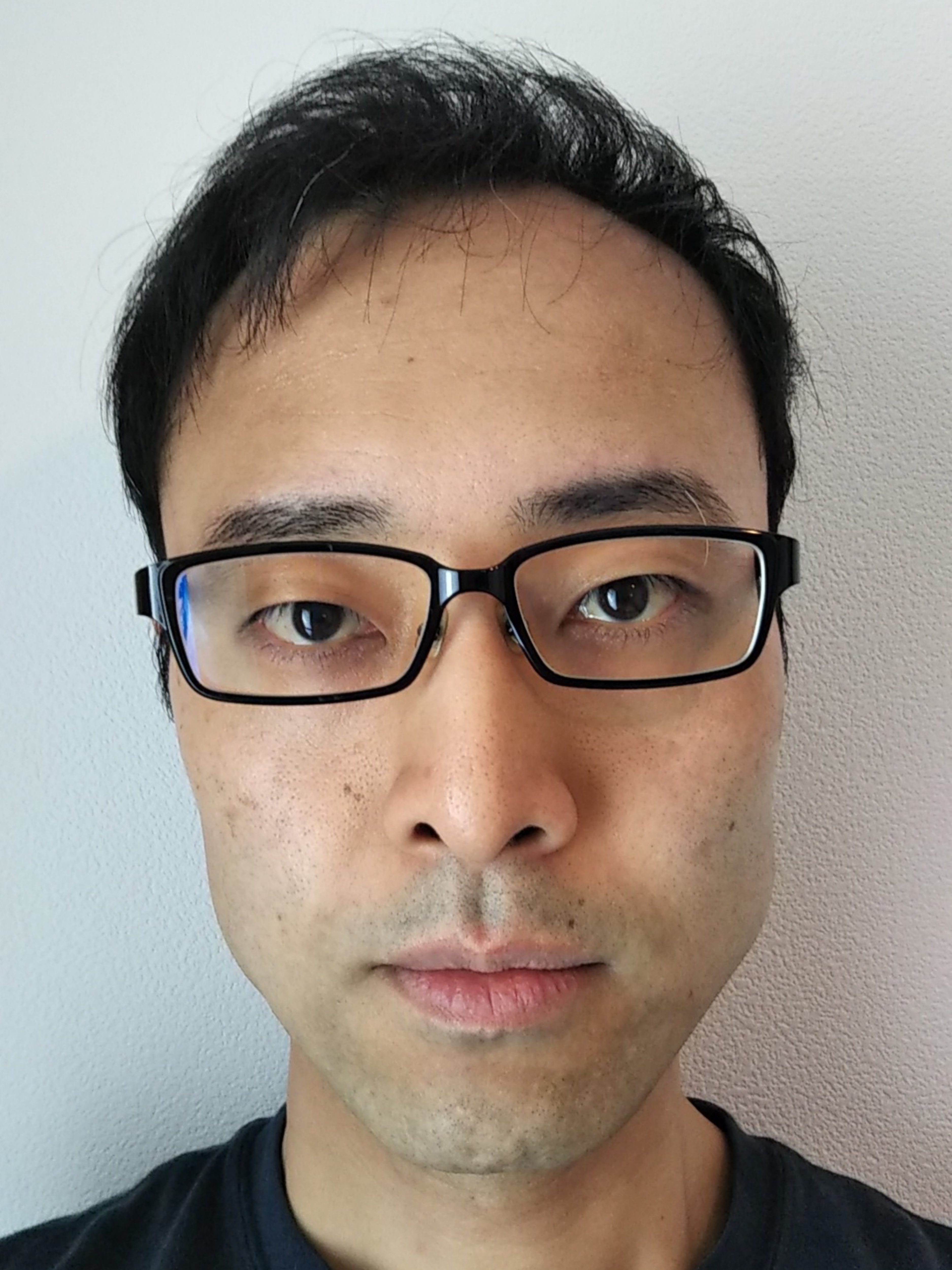]{Kenji Hirose}{
received the BA and MA degrees in human sciences, and the PhD degree in literature from Hokkaido University in 2003, 2005 and 2012, respectively. He is currently a postdoctoral researcher at Hokkaido University. He is also a visiting scholar at Hiroshima University. His research interests include the vividness of mental visual imagery and its relationship to the interoception, and the sense of reality and truthiness in theater. He is a member of JPA, JIA, and JSCP.
}

\profile[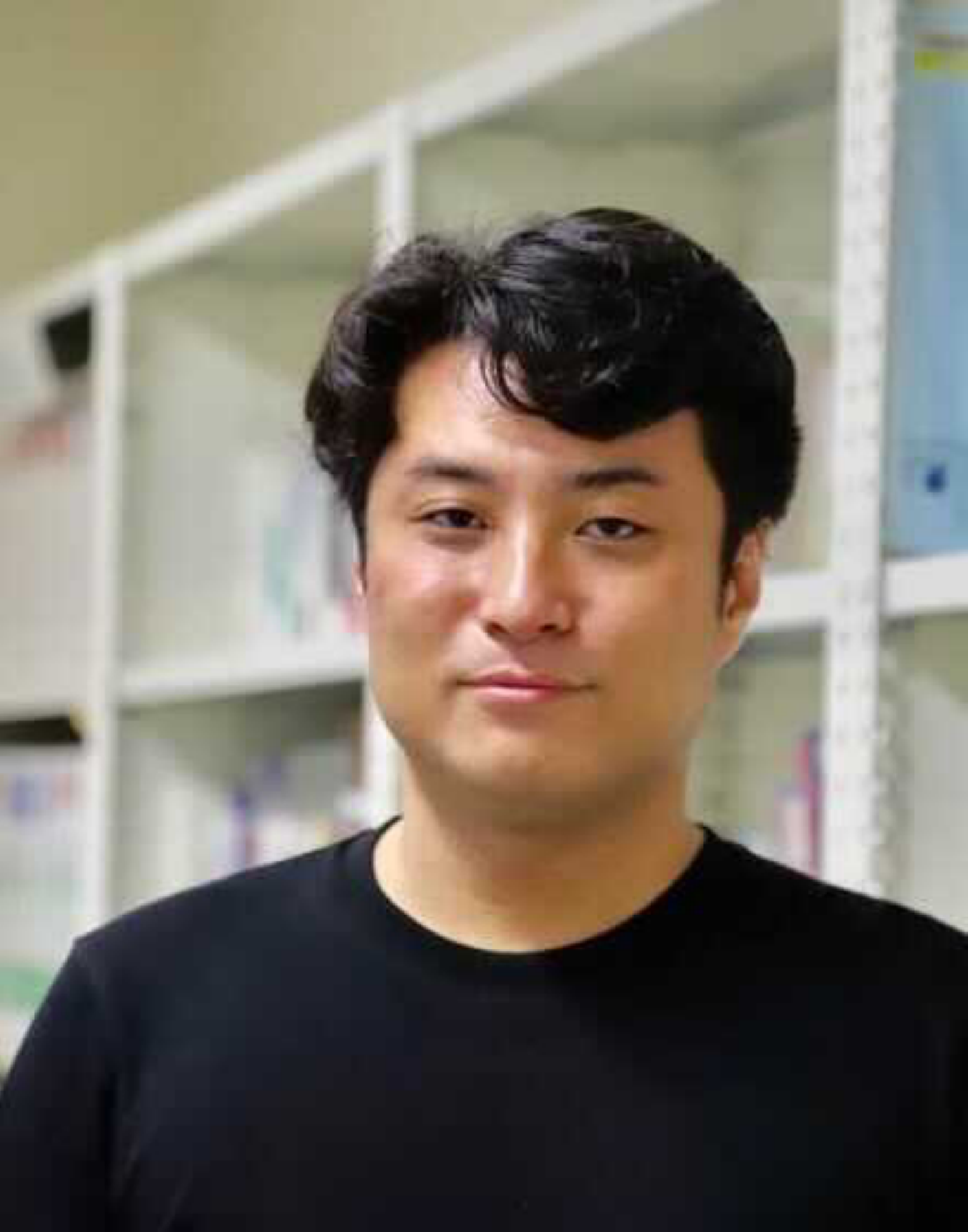]{Takayoshi Kitamura}{
received the BA degree from Kagoshima University, Japan and ME and PhD degrees from Kyoto University, Japan, in 2009, 2012, and 2018, respectively. In 2015, he joined Ritsumeikan University, where he was an Assistant of the College of Information Science and Engineering. In 2021, he joined Kagawa University, where he is currently the associate professor of media design studio. His research interests include communication design, information design, sound scape design, and so on. He is a member of the Human Interface Society.
}

\end{document}